\documentclass[journal]{IEEEtran}
\usepackage{cite}

\ifCLASSINFOpdf
\else
\fi
\usepackage{graphicx}
\usepackage{caption}
\usepackage{capt-of}
\usepackage{balance}
\usepackage{subfigure}
\usepackage{fixltx2e}
\usepackage{setspace}
\usepackage{amsmath}
\usepackage{hyperref}
\usepackage{array}
\usepackage{amssymb}
\usepackage{breqn}
\usepackage[dvipsnames]{xcolor}
\begin{document}

\title{Real-Time and Incremental Learning Enabled User-centric Myoelectric  Prosthesis System}

\author{Sidharth~Pancholi,~\IEEEmembership{ Member,~IEEE, } Amit M. Joshi,~\IEEEmembership{Senior Member,~IEEE, } Deepak Joshi,~\IEEEmembership{ Member,~IEEE, and}   Bradley S. Duerstock ~\IEEEmembership{Member,~IEEE}
		
\thanks{This work was supported in part by Bionics and Intelligence Lab, Jaipur, India. and Ministry of Human Resource and Development, Govt. of India}
 \thanks{Sidharth Pancholi, is with Weldon School of Biomedical Engineering, Purdue University, USA (e-mail: s.pancholi@ieee.org)}
\thanks{Amit M. Joshi was with the Department of Electronics and Communication Engineering, Malaviya National Institute of Technology, Jaipur, Rajasthan,India e-mail: (amjoshi.ece@mnit.ac.in)}

\thanks{Deepak Joshi is with Centre for Biomedical Engineering, Indian Institute of Technology Delhi, New Delhi 110016, India and the Department of Biomedical Engineering, All India Institute of Medical Sciences, New Delhi 110029, India e-mail: (Deepak.Joshi@cbme.iitd.ac.in
)}

\thanks{B. S. Duerstock is with Weldon
School of Biomedical Engineering and School of Industrial Engineering , Purdue University, West Lafayette, IN
47907 USA e-mail: (bsd@purdue.edu)}
 \thanks{Manuscript received XXXXXX; revised August XXXXXXXX}}

\maketitle

\begin{abstract}
The research landscape in the field of EMG signal-based bionics and prosthesis has witnessed significant growth in the last decade. Traditional myoelectric prosthesis devices, although functional, are limited by a fixed set of arm motions, and the development of innovative control mechanisms for prostheses is a time-consuming and lab-intensive process. Addressing this gap, our work introduces a cutting-edge upper limb prosthetic application system employing real-time training and incremental learning techniques. The system is equipped with the ADS1298 analog front end (AFE) and a 32-bit Arm Cortex-M4 processor for digital signal processing (DSP). Crucially, our system caters to both intact and amputated subjects. We leverage time-derivative moment-based features for effective pattern classification. Initially trained for four classes via online training, the system dynamically adapts to user demands, incrementing the number of classes to eleven. Remarkably, the system demonstrated exceptional performance, achieving a flawless 100\% completion rate for both healthy and amputated subjects in the case of four considered motions. Even as the number of classes increased to eleven, the system maintained robust performance, boasting completion rates of 94.33\% and 92\% for healthy individuals and amputees, respectively. Furthermore, we rigorously evaluated the system's motion efficacy for all subjects. Our results revealed outstanding efficacy rates, with intact subjects achieving an efficacy rate of 91.23\% and amputated subjects demonstrating an efficacy rate of 88.64\%. The findings underscore the system's effectiveness and potential, positioning our work as a pioneering contribution in the domain of upper limb prosthetics. This research is poised for acceptance in premier academic and scientific forums, promising to make a significant impact within the research community.

\end{abstract}
	

\IEEEpeerreviewmaketitle

\section{Introduction}
\IEEEPARstart The Electromyography (EMG) signal serves as a crucial control signal in various applications such as human assistive devices \cite{farina2014extraction, pancholi2023use}, human-robot interfaces (HRI) \cite{chandra2018muscle}, and prosthesis applications \cite{li2022new}. Pattern recognition, employing classification algorithms, is commonly utilized to interpret user intentions. Despite numerous studies on EMG Pattern Recognition (EMG-PR) system development, existing approaches exhibit suboptimal performance and struggle with online training for multiple classes. Additionally, conventional myoelectric prosthesis devices feature fixed classes, hindering swift adaptation and requiring laborious, lab-centric processes for developing new control mechanisms \cite{nsugbe2021brain, unanyan2021design, pancholi2023advancing}.

To enhance prosthetic performance in terms of class manipulation and extension, a user-centric approach is imperative. Users should have the capability to autonomously extend and add classes as per their requirements.

The EMG-PR system entails the extraction of relevant parameters or features from EMG signals, classifying them based on distinct classes. Feature extraction methods encompass time domain (e.g., mean absolute value (MAV), waveform length (WL), zero crossing (ZC), slope sign change (SSC)), frequency domain (e.g., mean frequency, peak frequency power), and time-frequency domain (e.g., wavelet transform, stockwell transform). While time domain features, such as MAV, WL, ZC, and SSC, generally outperform frequency domain features, their efficacy diminishes during homogeneous activities. Frequency domain features exhibit poor pattern classification performance \cite{kirby2023time}.

Phinyomark et al.\cite{Phinyomark2013} introduced modified features, namely modified median frequency (MMDF) and modified mean frequency (MMNF), enhancing the robustness and classification performance of EMG signals. Techniques like min-max normalization and PCA-based feature extraction have been employed to augment real-time performance in EMG-PR prostheses. Wang et al. proposed a novel wavelet method for precise hand grasp force motion classification. Features derived from wavelet packet decomposition (WPD) improve accuracy in homogeneous activity classification but introduce heightened computational complexity. Time-derivative moment-based features offer a balanced compromise between time and frequency, presenting a less computationally complex option for classifying homogeneous upper limb motions.


In the realm of EMG signal analysis, the utilization of wavelet transform has been pivotal, particularly in two predominant forms: continuous wavelet transform (CWT) and discrete wavelet transform (DWT). Notably, the efficacy and consistency of DWT for feature extraction from EMG signals surpass those derived from CWT. Despite the acknowledged additional processing time associated with DWT, its superiority remains evident \cite{Phinyomark2012, pancholi2022source}.

An innovative approach by Wang et al. \cite{wang2017design} introduced a method for capturing hand grasp force motion with precision from surface electromyography (sEMG) values, employing the wavelet method. Additionally, features based on wavelet packet decomposition (WPD) have exhibited enhanced classification accuracy for homogeneous activities. However, it's crucial to note that these features entail a trade-off, as they are accompanied by high computational complexity \cite{duan2016semg, pancholi2022dlpr}.

In a parallel study by Pancholi et al. \cite{pancholi2019time}, a novel approach was proposed for the classification of homogeneous upper limb motions. This method involved time-derivative moment-based features strategically leveraging information from both time and frequency domains. The result is a balanced compromise, maintaining effective classification performance while reducing computational complexity. This research diversification underscores the dynamic landscape of feature extraction techniques in EMG signal processing \cite{pancholi2019time}.

\begin{figure*}[htbp]
	\centering
	\includegraphics[width=16cm,height=9cm]{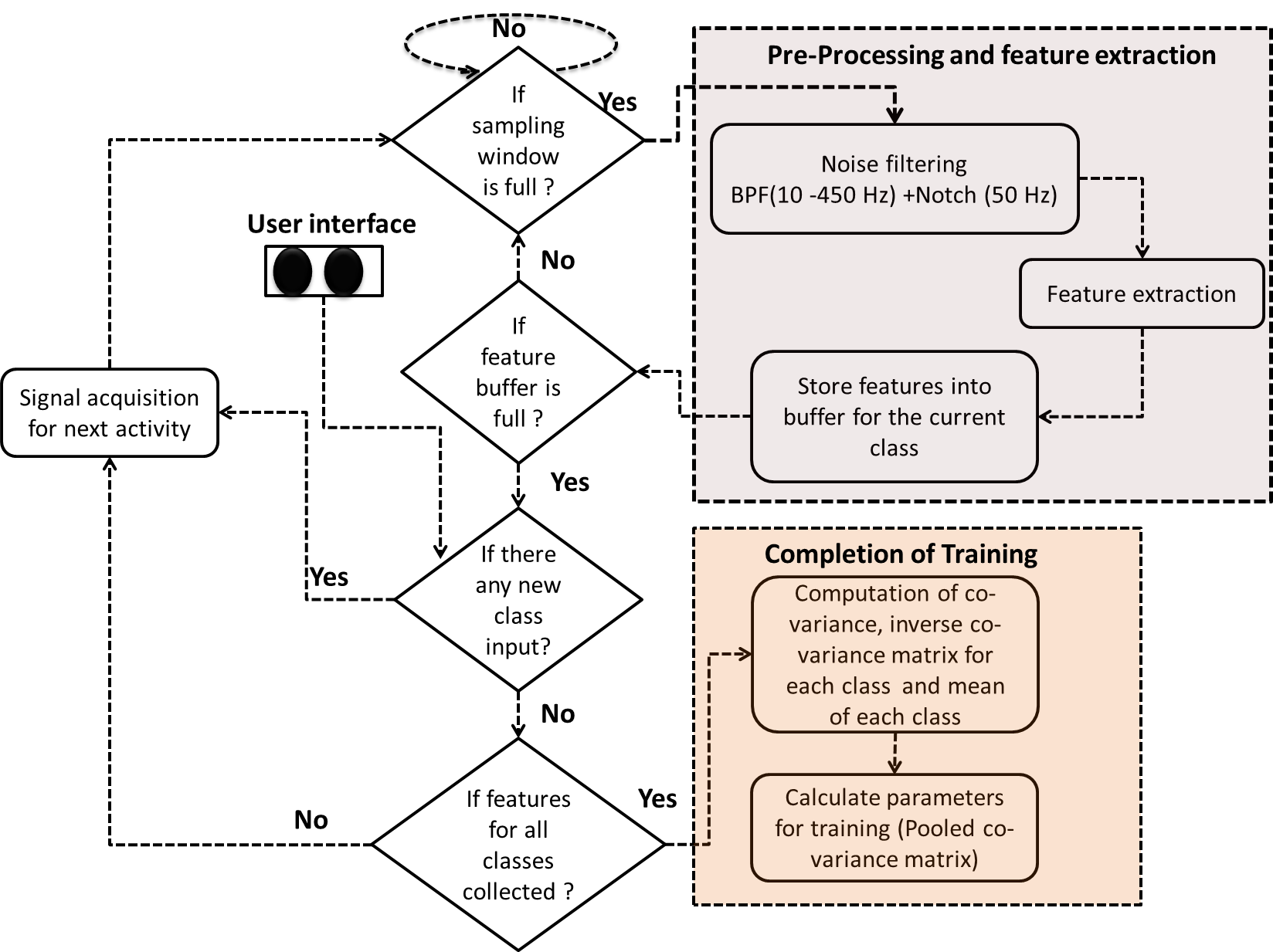}
	\caption{Flow chart of proposed algorithm for incremental learning using discriminant analysis }
	\label{flowchart}
\end{figure*}

Various classification algorithms, including LDA (linear discriminant analysis), QDA (quadratic discriminant analysis), SVM (support vector machine), k-NN (k nearest neighbor), and ANN (artificial neural networks), have found extensive application in EMG-based prosthetic systems \cite{samuel2018pattern, subasi2013classification, benatti2015versatile}. The existing literature features numerous EMG-based systems for both upper and lower limbs, often involving offline data processing on laboratory computers \cite{lu2017real, vijayvargiya2022hardware, pancholi2020advanced}. Notable advancements include the development of portable and wireless EMG signal acquisition systems \cite{pancholi2018portable, pancholi2019electromyography}. To enhance pattern recognition rates, sensor fusion techniques such as EMG-accelerometer, EMG-NIRs (near-infrared), and EMG-ultrasound waves have been suggested \cite{lu2017real, he2018wrist, pancholi_improve, pancholi2021intelligent}. However, these techniques often suffer from higher power consumption and increased system complexity.

The challenge of introducing new activities or classes has been addressed by researchers \cite{xu2018new, Pancholi2019}, proposing an incremental learning-based system designed for seamless class extension \cite{duan2018classification}. Incremental learning, an emerging machine learning technique, preserves existing knowledge, enabling easy integration of modifications based on new data. This approach eliminates the need for retraining the entire system, reducing memory consumption and training time. The adaptability of incremental learning extends beyond development, allowing systems to dynamically respond to user habits and conditions.

In the domain of EMG-based prosthetics, efforts have been concentrated on achieving simultaneous control of multiple degrees of freedom (DoF) \cite{krasoulis2020myoelectric}. High-Density EMG (HD-EMG) signal-based systems have demonstrated exceptional performance \cite{rahimi2018efficient}.

A novel incremental learning-based system employing Wavelet Neural Networks (WNN) has been introduced \cite{duan2016recognizing}. This system excels in continuous learning, enhancing classification performance for up to six hand gestures among amputated subjects. While more intricate than traditional approaches, the WNN proves effective in intricate gesture classification.

To enhance system efficiency, a hybrid approach combining Linear Discriminant Analysis (LDA) projection and Multi-Layer Perceptron Classification (MLP) has been proposed \cite{raurale2020real}. This combination demonstrates fast and efficient classification across a spectrum of activities.

For real-time system development, a comparative study between Force Myography (FMG) and EMG has been conducted \cite{belyea2019fmg}. Furthermore, the integration of real-time hardware with Internet of Things (IoT) technology has enabled robust EMG-based pattern recognition systems \cite{zanghieri2019robust}. Ongoing developments explore the potential of high-density EMG signals for enhanced feature extraction, paving the way for improved myoelectric prostheses \cite{clarke2020deep}.

Novel initiatives include the design of a Visual Feedback System for trans-humeral amputees \cite{tiwari2021design}, techniques for prosthesis generalization \cite{gulati2021toward}, and comprehensive control experiments \cite{ortiz2014real}.

	
\begin{figure*}[htbp]
	\centering
	\includegraphics[width=14cm,height=7cm]{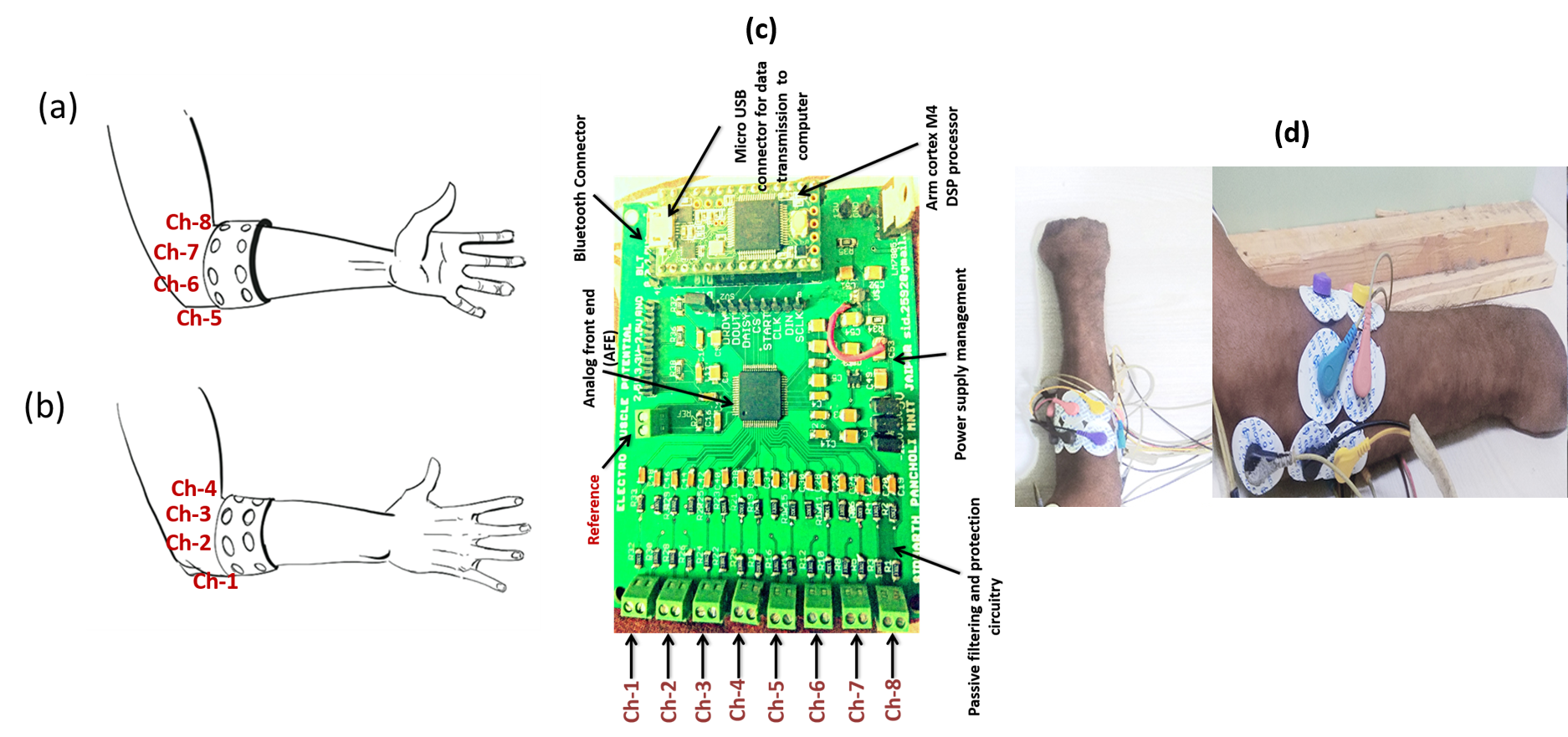}
	\caption{Experiment and system depiction: (a) Anterior view (b) Posterior view  with the proposed system (c) proposed embedded platform (d) depiction of the amputated arm with electrodes}
	\label{arm}
\end{figure*}

\begin{figure*}[htbp]
	\centering
	\includegraphics[width=14cm,height=9cm]{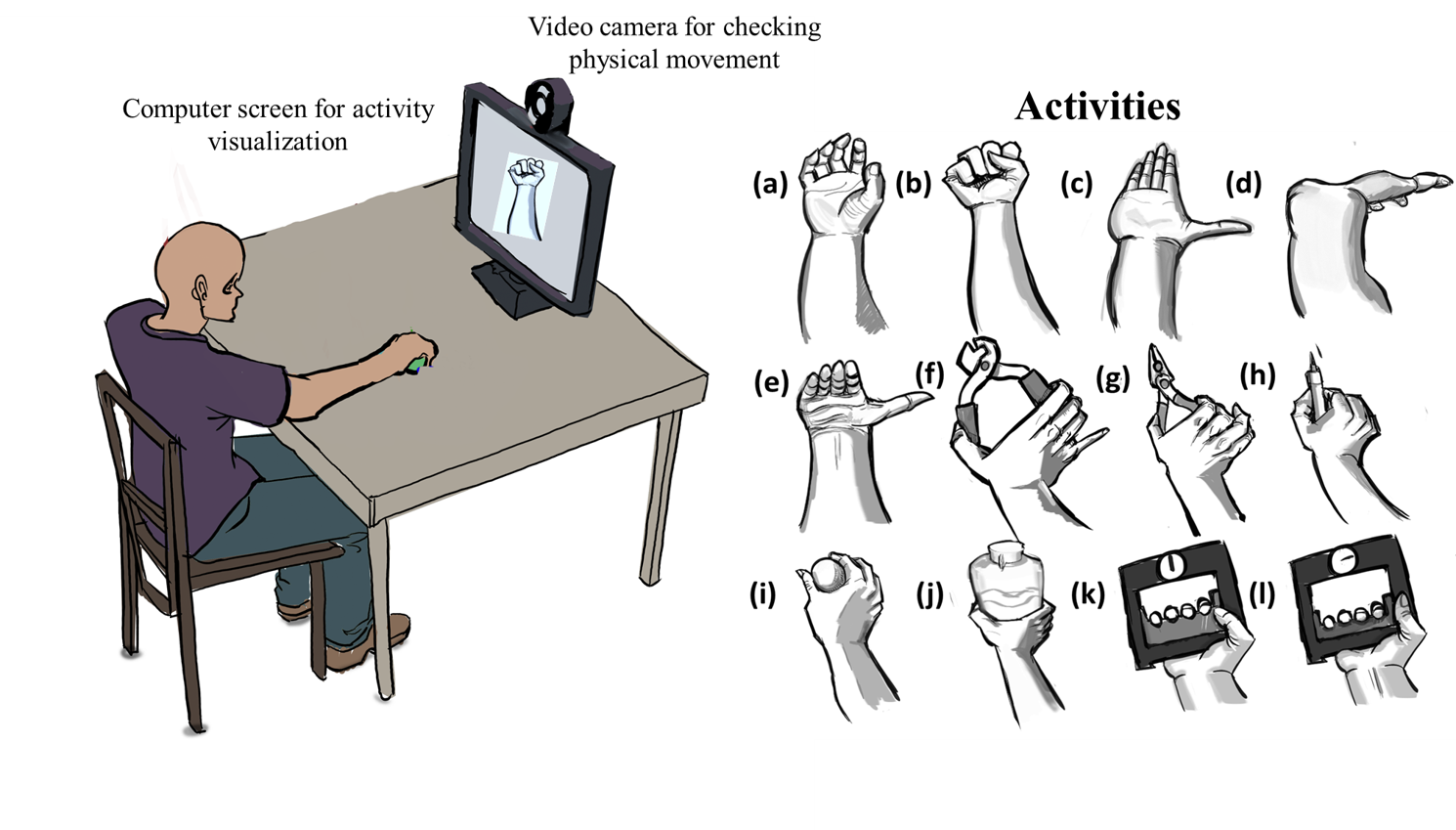}
	\caption{(a) Rest (b) Hand close (c) Hand open (d) Wrist extension (e) Wrist flexion (f) Cutter grasp (g) pliers grasp (h) Screw grasp (i) Quadrupod grasp (j) Large diameter grasp (k) Normal parallel extension grasp (l) Forced parallel extension grasp }
	\label{act}
\end{figure*}


	Considering the benefits and drawbacks of the research work in the literature. An EMG-PR system adapted to a new class (hand motion not used during model training) with minimal computational complexity has been developed. A system like this would be more useful to amputees and make it easier for people to accept smart prostheses. The main contributions of this work are as follows: 
	
	\begin{enumerate}
	\item A new feature extraction approach has been implemented and tested in real-time for effective EMG-PR. 
	\item An incremental learning-based EMG-PR system has been developed for upper limb amputees. 
	\item The system can classify up to 11 upper limb motions with reasonable classification accuracy on amputated subjects. 
	\item  The pattern completion performance is evaluated using the motion completion test and motion efficacy test. 
	\end{enumerate}

The paper flow is as follows: Section II concerns the design of the system and the acquisition of EMG signals. Section III explains the method for extraction of EMG features extraction. Results and discussions are listed in Section IV. Subsequently, the conclusion is drawn from Section V.

\section{Methodology}
\subsection{System Architecture}
The system is equipped with an ADS1298 serving as an Analog Front End (AFE), featuring an 8-channel differential feed with an advanced band and adjustable strap for non-invasive EMG signal acquisition through metallic $Ag-AgCl$ patches. To ensure high input impedance, protection circuits with a $200M Ohm$ resistance are integrated before each channel's input. Power management is facilitated by a power control circuit, drawing power from a $3.7V$ lithium-ion battery.

For digital signal processing, a DSP processor with a $32$-bit architecture and an ARM Cortex-M4 core operates at a clock speed of 150 MHz, utilizing the built-in Floating Point Unit (FPU). The data acquisition frequency is set at $1000SPS$ (samples per second). To eliminate DC and power line noises, a $3^{rd}$ order IIR Band Pass Filter (BPF) spanning from 10 Hz to $500Hz$ is implemented, cascaded with a $2^{nd}$ order $50Hz$ notch filter. MATLAB 2015a software is employed for extracting filter coefficients.

The algorithm implemented is an incremental learning-based Linear Discriminant Analysis (LDA) classifier, allowing user input for class extension. The flowchart of this algorithm is depicted in Fig. \ref{flowchart}. Raw EMG signals from the 8-channel differential inputs are digitized through the AFE and stored in a circular buffer. Subsequently, feature extraction is performed, and the proposed algorithm utilizes these features for activity-specific training.

The overall system architecture is illustrated in Fig. \ref{arm}, showcasing the integration of components for effective EMG signal processing and pattern recognition. This sophisticated setup ensures high-performance real-time processing and adaptability to user-defined classes, enhancing the system's technical capabilities for diverse applications.
	
\subsection{EMG Signal Acquisition and Experimental Setup}

\begin{enumerate}

\item 
A total of 8 electrodes have been employed around the forearm of the residual part of the limb for EMG signal acquisition as depicted in Fig.\ref{arm}. Similar electrode positions have been considered for healthy subjects after cleaning the electrode position with 20\% (v/v) alcohol. 
\item
A total of 8 subjects including 5 healthy (intact) male subjects (age: 22$\pm$4.2) and 3 amputees have been recruited for this study. The information of the amputated subjects is given in Table \ref{amputee_info}.
			
\begin{table}[h]
\caption{Information about amputees}
\begin{tabular}{llllll}
\hline
\hline
\begin{tabular}[c]{@{}l@{}}Subject\\  ID\end{tabular} & Age & \begin{tabular}[c]{@{}l@{}}Height\\  (cm)\end{tabular} & \begin{tabular}[c]{@{}l@{}}Weight\\  (kg)\end{tabular} & Gender & Reason \\ \hline
Amp-A                                                 & 27  & 170                                                    & 65                                                     & M      & Trauma \\
Amp-B                                                 & 60  & 172                                                    & 80                                                     & M      & Trauma \\
AMP-C                                                 & 40  & 168                                                    & 72                                                     & M      & Trauma \\ \hline \hline
\end{tabular}
\label{amputee_info}
\end{table}
		
\item A computer screen has been incorporated in the form of each subject for visual instruction. They have requested to sit comfortably in the chair.
		
\item  The experiment and other parts of the study were first explained to the subjects. It took approximately one hour, and then another half hour was allotted for primary training facilitation.

\item Initially, data for four classes are recorded in real-time, and an offline classification model is embedded in the system. Furthermore, the user can increase the number of classes by pressing the system's button until it reaches 11. This was done with a screen in front of the individual, and all activity that appeared had to be completed within 300 milliseconds for accurate performance as shown in Fig. \ref{act}. 

\item After each run, 2 minutes break has been given to avoid muscle fatigue. More detail has been elaborated on in the result section. 
		
\end{enumerate}

\section{ATDM Feature Extraction Scheme for EMG-PR}
In the development of our feature extraction methodology for EMG-PR systems, our goal was to establish a robust approach that incorporates both time and frequency information. Our thought process involved delving into the intricacies of EMG signal generation during muscle movement and devising features that capture relevant aspects of this process.

\textbf{Functional Derivative:}
\begin{align}
\mathcal{F}[D^{n}x[j]] &= \nu^{n}\mathcal{X}[\nu]
\end{align}

Here, $\mathcal{F}$ denotes the functional derivative, $D^{n}x[j]$ represents the $n^{th}$ derivative of the signal $x[j]$, and $\mathcal{X}[\nu]$ is the Fourier transform of the signal. This step is crucial for addressing frequency information using the Fourier transform.

\textbf{Power Spectral Moments:}
\begin{align}
\mu_{n} &= \sqrt{\sum_{\nu=0}^{L-1}\nu^{n}\mathcal{X}[\nu]}
\end{align}

Power spectral moments ($\mu_{n}$) are employed to capture features based on the power spectrum of the signal, where $\nu$ ranges from 0 to $L-1$.

\textbf{Parseval's Theorem:}
\begin{align}
\sum_{j=0}^{L-1}\left | x\left [ j \right ] \right |^{2} &= \sum_{\nu=0}^{L-1}P[\nu]
\end{align}

Parseval's theorem ensures consistency between time and frequency domains, where $\left | x\left [ j \right ] \right |^{2}$ is the squared magnitude of the signal and $P[\nu]$ is the power spectrum.

\textbf{Feature Extraction:}
\begin{align}
\mu_{0} &= \sqrt{\sum_{j=0}^{L-1}(x[j])^{2}} \\
\mu_{2} &= \sqrt{\sum_{j=0}^{L-1}(Dx[j])^{2}} \\
\mu_{4} &= \sqrt{\sum_{j=0}^{L-1}(D^2 x[j])^{2}}
\end{align}

The feature extraction process involves calculating the $0^{th}$ order moment ($\mu_{0}$) and the $2^{nd}$ order time derivative moment ($\mu_{2}$).

\textbf{Number of Peaks (NPs) and Zero-Crossings (ZCs):}
\begin{align}
NPs &= \sqrt{\frac{\mu_{4}}{\mu_{2}}} \\
ZCs &= \sqrt{\frac{\mu_{2}}{\mu_{0}}}
\end{align}

These moments are then used to define the number of peaks (NPs) and the number of zero-crossings (ZCs) in the stochastic process.

\textbf{Reduced Complexity:}
\begin{align}
NPs &= \sigma \\
ZCs &= \theta
\end{align}

To simplify computations, square versions of NPs and ZCs are introduced, denoted as $\sigma$ and $\theta$.

\textbf{Feature Formulation:}
\begin{align}
PAP &= \frac{\mu_{0}}{\sigma} \\
ZCAP &= \frac{\mu_{0}}{\theta} \\
MWL &= \sum_{0}^{L-1}s_{i}^{'}-s_{i-1}^{'} \\
DBM &= \mu_{0} - \mu_{2}
\end{align}

Here, $PAP$ represents Peak Average Power, $ZCAP$ is Zero Crossing Average Power, $MWL$ is Modified Waveform Length, and $DBM$ is the Difference between Moments. These features are formulated based on the derived moments, aiming to provide a reliable and comprehensive approach for decoding human intent from EMG signals.

\subsection{LDA based Incremental Learning}
Incremental learning is a method through which the model is trained incrementally using new data. The key advantage of incremental learning is that the generated model dynamically adapts to new patterns in the current model parameters. The linear discriminant analysis-based classifier has the inherent advantage of low operational memory overheads \cite{chu2015incremental}. 
This is due to its principle of operation where a single pooled co-variance matrix gets utilized irrespective of the number of presence of the classes. When this technique is used, the following benefits can be obtained:

\begin{enumerate}
    \item It is capable of learning new information from new samples.
    \item It is not essential to process all previously stored information when upgrading. It also saves previously acquired knowledge.
    \item It can introduce additional classes as per the user's demand.
\end{enumerate}

The Discriminant Analysis (DA) models the probability density function (PDF) of each class $k$ as follows:

\begin{align}
f_k(x) = \frac{1}{(2\pi)^{d/2}}\frac{1}{\left | \xi_k \right |^{1/2}}\exp\left(-\frac{(x-\mu_k)^T\xi_k^{-1}(x-\mu_k)}{2}\right)\pi_k
\end{align}

In this equation, $x$ (input) is a vector in $\mathbb{R}^d$, $\mu_k$ (mean) is a vector in $\mathbb{R}^d$, $\xi_k$ is the covariance matrix in $\mathbb{R}^{d\times d}$, and $\left | . \right |$ denotes the determinant of a matrix. The parameter $\pi_k$ represents the prior probability of class $k$.

Taking the natural logarithm of the PDF, we obtain the log-likelihood function $\alpha_k(x)$ for class $k$:

\begin{align}
\alpha_k(x) = -\frac{1}{2}\ln(\left | \xi_k \right |) - \frac{1}{2}(x-\mu_k)^T\xi_k^{-1}(x-\mu_k) + \ln(\pi_k)
\end{align}

Linear Discriminant Analysis (LDA) assumes that the covariance matrix is the same for all classes, denoted as $\xi$. Therefore, the log-likelihood function for class $k$ simplifies to:

\begin{align}
\alpha_k(x) = \mu_k^T\xi^{-1}x - \frac{1}{2} \mu_k^T\xi^{-1}\mu_k + \ln(\pi_k)
\end{align}

To classify a new instance $x$, we choose the class with the highest log-likelihood:

\begin{align}
\widehat{C} = \underset{k}{\text{argmax}} \ \alpha_k(x)
\end{align}

Suppose a new class $T$ is introduced in the existing trained model. The new covariance matrix is denoted as $\xi_{\text{new}}$ and computed by adding the new class's covariance matrix $\xi_T$ to the existing pooled covariance matrix $\xi$:

\begin{align}
\xi_{\text{new}} = \xi + \xi_T
\end{align}

These equations describe the Gaussian Discriminant Analysis and Linear Discriminant Analysis along with how to update the pooled covariance matrix when a new class is introduced.

\begin{figure*}[htbp]
	\centering
	\subfigure[]{\includegraphics[width=0.36\textwidth,height=0.28\textwidth]{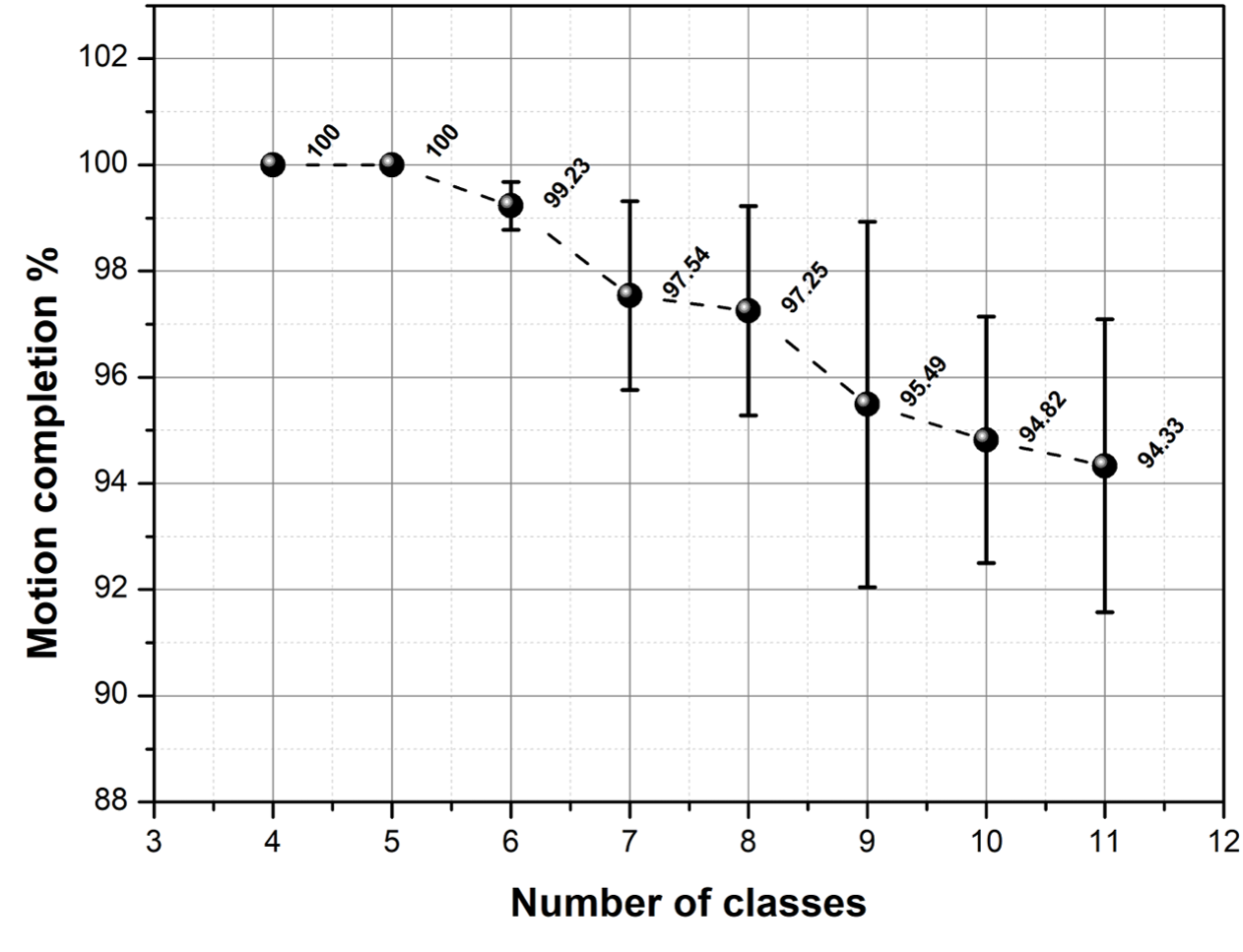}}
	\subfigure[]{\includegraphics[width=0.36\textwidth,height=0.28\textwidth]{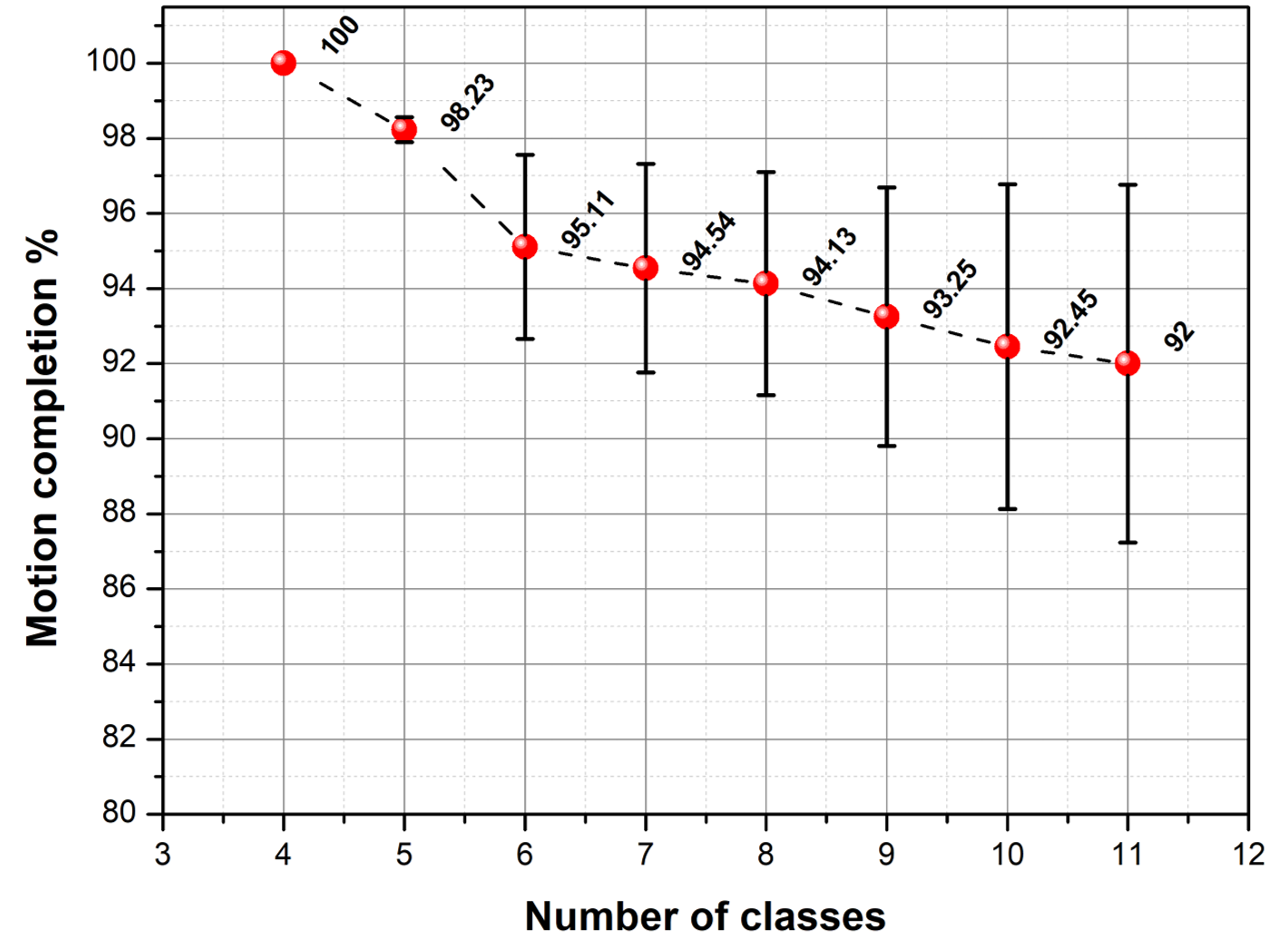}}
	\subfigure[]{\includegraphics[width=0.36\textwidth,height=0.28\textwidth]{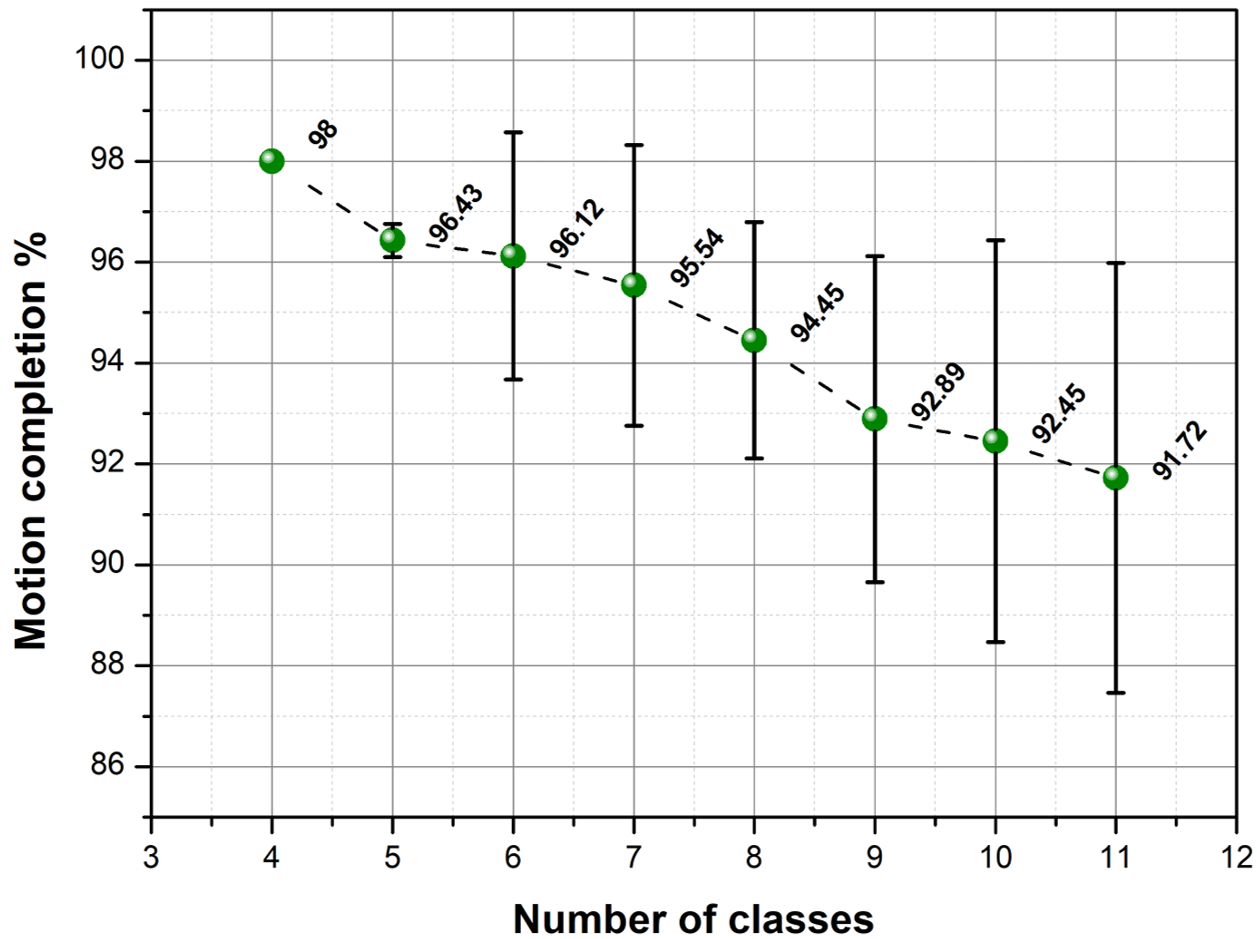}}
	\subfigure[]{\includegraphics[width=0.36\textwidth,height=0.28\textwidth]{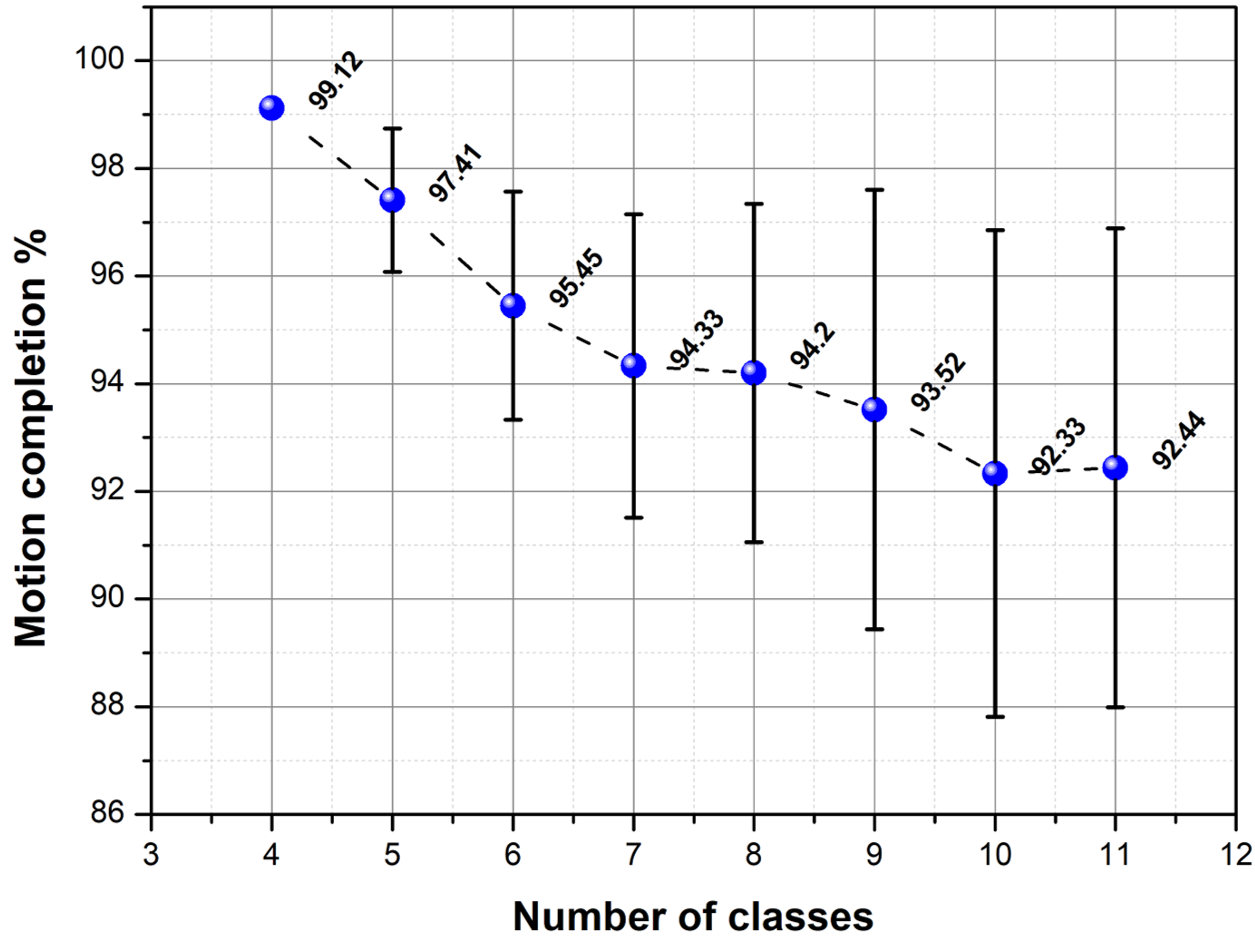}}
	\caption{Real-time incremental learning performance: (a)  Healthy subjects,  (b) Amputee-A, (c) Amputee-B, (d) Amputee-C }
	\label{result}
\end{figure*}

These four main sub-processes, signal acquisition, segmentation, feature extraction, and classification (testing), must be performed with a latency of 200-300 ms \cite{Smith2011} to satisfy the real-time device constraints. In this work, EMG data was segmented into a series of windows of 200 ms with a window shift of 75 ms. This helps to reduce the non-stationary factor of the signal and makes the signal quasi-stationary.

\begin{table*}[htbp]
\caption{Comparison of the proposed work with recent literature work}
\footnotesize{
\begin{tabular}{llllllllll} 

\hline \hline
\begin{tabular}[c]{@{}l@{}}Lit. work $\Rightarrow$ \\ \\ Parameters $\Downarrow$ \end{tabular} & \begin{tabular}[c]{@{}l@{}}Duan et al.\\  (2017) \cite{duan2016recognizing}\end{tabular} & \begin{tabular}[c]{@{}l@{}}Pancholi et al.\\   (2019) \cite{pancholi2019electromyography}\end{tabular} & \begin{tabular}[c]{@{}l@{}}Yu et al.\\     (2019) \cite{yu2019novel}\end{tabular} & \begin{tabular}[c]{@{}l@{}}Raurale et al.\\       (2020) \cite{raurale2020real}\end{tabular} & \begin{tabular}[c]{@{}l@{}}Tam et al.\\      (2021) \cite{tam2021intuitive}\end{tabular} & \begin{tabular}[c]{@{}l@{}}Moin et al.\\      (2021) \cite{moin2021wearable}\end{tabular} & Proposed \\ \hline \hline
Subject & Healthy & Healthy/Amp. & Healthy & Healthy/Amp. & Healthy & Healthy/Amp. & Healthy \\
No. of classes & 6 & 5 & 12 & 11+rest & 6 & 13/21 &  \\
Classifier & WNN & LDA & LDA & \begin{tabular}[c]{@{}l@{}} LDA with MLP\end{tabular} & CNN & \begin{tabular}[c]{@{}l@{}}Neuro-inspired \\ hyperdimensional \\ computing \\ algorithm\end{tabular} & LDA \\
Feature Techniques & RMS & \begin{tabular}[c]{@{}l@{}}RMS, SSC,\\ ZC, WL\end{tabular} & MPT & Time domain  & Hypervector & ATDM &  \\
Classification accuracy & 92.17 & 94.14 & 86.61 & 99.30 & 93.43 & 97.12 / 92.87 &  \\
Real-time & No & Yes & Yes & Yes & Yes & Yes & Yes \\
Embedded & No & Yes & No & Yes  & No & Yes & Yes \\
Incremental & Yes & No & No & No & No & Yes & Yes \\
Amputees & No & No & No & No & No & Yes & No \\ \hline \hline
\end{tabular}}
\label{comp}
\end{table*}

\section{Performance evaluation}
In order to evaluate the performance, the following metrics are considered:

\begin{itemize}

   \item \textbf{Motion Completion (MC)}: This metric assesses the performance of the proposed technique in real-time scenarios. The algorithm predicts target motions, and the processed feature vector is considered successful if the predictions are correct. The effectiveness of the classifier is calculated as the ratio of correct predictions to total predictions over all windows ($N$) and multiplied by 100. A classification score of 100\% is achieved if the subject successfully achieves all motions, while a score of 0\% denotes no target motion achieved. The MC metric is represented by Equation (\ref{MC}).

\item\textbf{Motion Efficacy}: The motion efficacy metric combines the classifier prediction and the speed of proportional control, as shown in Equation (\ref{eq:eff}). This metric takes into account both the classification success rate and the volitional speed used to control the prosthetic arm. A score of 100\% indicates that the subject controlled the prosthetic arm correctly and smoothly with the appropriate speed. Conversely, a score of 0\% suggests poor control over the prosthetic limb and a failure to achieve the target motion.

\begin{align}
prop_{n}^{pro} = prop_n \times est_{n}^{pro}
\label{prop}
\end{align}

\begin{align}
MC = \left( \frac{\sum_{n=1}^{N} est_n^{pro}}{N} \right) \times 100
\label{MC}
\end{align}

Here, $prop_n$ represents the volitional speed used to control the prosthetic limb.

\begin{align}
{	est_{pro}^n=\begin{cases}
		1 & \text{ if } subject \quad achieves\quad  the \quad motion \\ 
		0 & \text{ if }  no \quad  motion \quad or\quad wrong \quad motion\\ 
		\end{cases}}
	\label{est}
\end{align}

\begin{align}
eff^{motion}=\left [ \sum_{1}^{n} \frac{prop_{n}^{pro}}{prop_n}\right ]
\label{eff}
\end{align}
\end{itemize}

\section{Result and Discussion}

\subsection{Motion completion performance}	To evaluate the accuracy of the  proposed system,  the subject is asked to repeat each motion 30 times. In this work, four classes have been taken into initial consideration, which later on gets incremented to 11 classes and testing  has been performed. The accuracy of the classifier has been evaluated initially for four classes, and later on, accuracy for individual class increments has been analysed and is depicted in Fig. \ref{result} (a) for intact subjects. The classification rate patterns are summarized for each amputee in Fig. \ref{result} (b), (c), and (d). The initial motion completion rate is found to be 100\% which stays as such when the number of classes is increased to 5 for healthy subjects. Additional statistics show that when the number of classes is raised to 11, the completion rates for both healthy and amputated participants are 94.33\% and 92\%, respectively.

\begin{figure}[htbp]
	\centering
	\includegraphics[width=7.5cm,height=6cm]{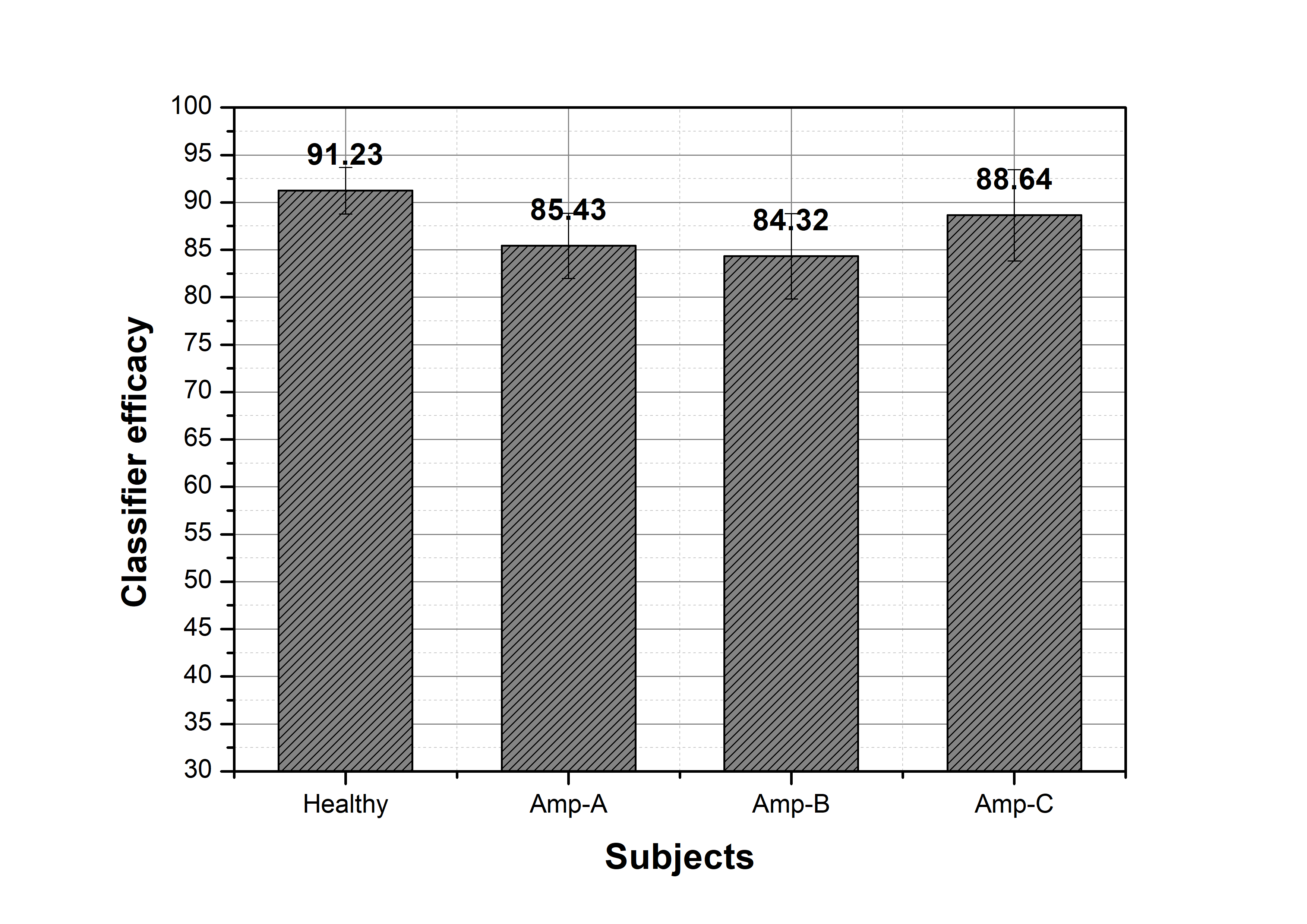}
	\caption{motion efficacy of test result}
	\label{effy}
\end{figure}

\subsection{Motion efficacy}

The motion efficacy of all subjects including healthy and amputees is shown in Fig. \ref{effy}. This represents that the higher  and smooth classification performance within the time frame and speed. The healthy subject shows above 91.23\% efficacy rate. The highest motion efficacy rate 88.64\% has been obtained when amputated subjects are considered.\par
The comparison with state of art systems is summarized in Table \ref{comp}. Several studies on myoelectric prostheses were exclusively used intact individuals only \cite{liu2016virtual}. In the research \cite{vidovic2015improving},  the performance of the proposed technique was evaluated using online testing, but the system was not realized on an embedded platform. In the studies \cite{turlapaty2019feature, yu2019novel, duan2016recognizing}, only off-line analysis was performed and no real-time adaptation for the system. Moreover, complex classification algorithms such as wavelet neural networks and convolution neural networks were adopted. However, the robustness of the system is required to be improved for real-time implementation.

In this work, versatile data sets including amputees have been used for system validation with real-time incremental learning on an embedded platform. 
Furthermore, the ANOVA (Analysis of Variance) test has been utilized to validate the real-time feasibility. This has been done in MATLAB R2018B.  The online evaluation gives more than 90\% p$>$ 0.96 motion completion rate  and  more than 90\% p$>$ 0.95 motion efficacy when metrics are considered for all the subjects. It is observed that there is a significant improvement in motion completion and motion efficacy performance with respect to TD features and time-frequency (wavelet)  at p$<$ 0.031.

\section{Conclusion}
The paper presents an incremental learning-based EMG-PR system that has been validated for intact and amputated subjects. The system shows enhanced performance over existing systems overcoming their limitations of off-line training and increased processed overheads with class extension. Time derivative moment-based features have been used due to their advantage of utilizing time and frequency domain information without any transformation. Moreover, the LDA classifier has ensured low computation complexity for better real-time performance.\par
	
The objective of this research is to improve the accuracy of this system and to conduct more complex and real-world experiments in order to transfer this research into a real prosthesis arm. For this reason, it has been decided to conduct a virtual reality experiment in order to better align the brain with activities using an amputated limb.

	\section{Acknowledgement}
	
The authors are thankful to the Ministry of Human Resource and Development, India, and Bionics and Intelligence Lab, Jaipur, India to provide facilities and funding for this work. The full version of the paper is available at open access database \cite{pancholi2021novel}.

	\bibliographystyle{IEEEtran}
	\bibliography{sid}
\end{document}